\documentclass[aps,prd,preprint,superscriptaddress,amsmath,amssymb,showpacs]{revtex4-1}
\usepackage{dcolumn}
\usepackage{graphicx}
\usepackage{float}
\usepackage{physics}
\usepackage[colorlinks=true,allcolors=blue]{hyperref}
\begin{document}
	\title{Non-extensive NJL model study of QCD phase structure with chiral imbalance and strong magnetic fields}
	
	\author{Xiang-Qiong Liu}
	\affiliation{College of Mathematics and Physics, China Three Gorges University, Yichang 443002, China}
	
	\author{Sheng-Qin Feng}
	\email{Corresponding author: fengsq@ctgu.edu.cn}
	\affiliation{College of Mathematics and Physics, China Three Gorges University, Yichang 443002, China}
	\affiliation{Center for Astronomy and Space Sciences and Institute of Modern Physics, China Three Gorges University, Yichang 443002, China}
	\affiliation{Key Laboratory of Quark and Lepton Physics (MOE) and Institute of Particle Physics,\\
		Central China Normal University, Wuhan 430079, China}

	\begin{abstract}
	Based on the two-flavor NJL model with Tsallis non-extensive statistics, this work explores the QCD phase structure and thermodynamic properties under strong magnetic fields and chiral imbalance. The Tsallis parameter $q$ captures non-equilibrium effects relevant to heavy-ion collisions. Key findings reveal that the critical temperature \(T_c\) decreases with increasing $q$, indicating that non-equilibrium conditions promote chiral symmetry restoration at lower temperatures. The chiral chemical potential $\mu_5 $ significantly alters the magnetic response, with a transition from magnetic catalysis to inverse magnetic catalysis under certain conditions. For $q > 1$, non-monotonic behavior of \(T_c\) with magnetic field $eB$  emerges. Pressure becomes anisotropic under strong $eB$, and the speed of sound exhibits a dip near \(T_c\), shifting to lower temperatures with larger $q$. These results highlight how non-extensive statistics, chiral imbalance, and magnetic fields collectively influence the QCD phase diagram and thermodynamic observables, offering insights for interpreting heavy-ion collision data.
	\end{abstract}
	
	\maketitle
	
	\section{Introduction}\label{sec:01_intro}
	The Nambu--Jona-Lasinio (NJL) model is an effective low-energy theory constructed from interacting chiral Dirac fermions \cite{rr1,rr2,rr3}. It plays a significant role in quantum chromodynamics (QCD), particularly in the non-perturbative regime where the large coupling constant renders perturbative methods inapplicable ~\cite{rr4,rr5}. Under extreme conditions such as strong magnetic fields, high temperatures, and high densities, quark matter exhibits rich and novel phase structures.
	
	In relativistic heavy-ion collisions conducted at facilities such as the Relativistic Heavy Ion Collider (RHIC) and the Large Hadron Collider (LHC), magnetic fields as strong as $eB\sim0.01 ~\textrm{GeV}^{2}$ and $eB\sim0.2 ~\textrm{GeV}^2$, respectively, can be generated. Although short-lived, the quark-gluon plasma (QGP) produced in such collisions exhibits a substantial medium response to magnetic fields, significantly delaying their decay \cite{rr6,rr7,rr8,rr9}. It is thus reasonable in certain contexts to assume the presence of a nearly constant external magnetic field. The introduction of strong magnetic fields enriches the QCD phase diagram and its thermodynamic properties, leading to emergent phenomena such as the chiral magnetic effect \cite{rr10,rr11,rr12,rr13}, magnetic catalysis (MC) in the vacuum \cite{rr14,rr15,rr16}, and inverse magnetic catalysis (IMC) near the chiral phase transition \cite{rr17,rr18,rr19,rr20,rr21,rr22}.
	
	The QCD vacuum contains non-trivial gluon configurations, such as instantons \cite{rr23} or sphalerons \cite{rr24,rr25} characterized by an integer topological winding number $Q_{\mathrm{w}} $. The high-temperature QGP created in heavy-ion collisions \cite{rr26} is expected to contain a high density of such topological excitations. These configurations can arise at any spacetime point within the QGP, locally altering the topological charge of the vacuum. Through the QCD axial anomaly, non-zero topological charge induces chiral imbalance, quantified by the chiral charge $N_{5}=N_{\mathrm{R}}-N_{\mathrm{L}}$, which is linked to the topological winding number of the gluon field via the Adler- Bell- Jackiw anomaly. This imbalance is commonly modeled by introducing a chiral chemical potential $\mu_5$, which becomes especially relevant in the presence of magnetic fields. The NJL model offers a powerful effective framework for studying chiral and deconfinement phase transitions in the context of non-central heavy-ion collisions, where strong magnetic fields and rotation are present \cite{rr27,rr28,rr29,rr30,rr31,rr32,rr33,rr34,rr35,rr36,rr37,rr38,rr39,rr40,rr41,rr42,rr43}. Its local four-fermion interaction preserves chiral symmetry, although gluonic degrees of freedom are integrated out. To account for topological effects and chiral imbalance arising from non-trivial gluon backgrounds, the chiral chemical potential $\mu_5$ is introduced within the QGP medium.
	
	In heavy-ion collisions at RHIC and LHC, the extremely short evolution timescale of the system keeps the QGP away from thermal equilibrium. This challenges the applicability of Boltzmann--Gibbs (B-G) statistics, which assumes local equilibrium, in describing phase transitions and critical phenomena \cite{rr44}. To address this, Constantino Tsallis introduced a non-extensive generalization of statistical mechanics in 1988 \cite{rr45,rr46,rr47}, incorporating a real parameter $q$ that quantifies the degree of non-extensivity. This framework effectively describes systems with long-range interactions, multi-scale correlations, and non-ergodic dynamics. In high-energy physics, Tsallis statistics has been particularly successful in capturing non-equilibrium and critical behaviors in strongly interacting systems \cite{rr44,rr48,rr49,rr50,rr51}, offering a valuable tool for exploring QGP evolution.
	
	In the framework of Tsallis non-extensive statistics, the introduction of the non-extensive parameter $q$ (which reduces to Boltzmann--Gibbs statistics when $q=1$) enables a more effective description of non-equilibrium behavior and critical phenomena in strongly interacting systems. Existing studies have indicated \cite{rr52,rr53} that the distribution function based on Tsallis entropy can effectively fit the transverse momentum spectra observed in high-energy collision experiments. In such descriptions, the parameter $q$ is typically chosen within the range of $1-1.2$. This characteristic is highly relevant in the context of the NJL model and studies involving the chiral chemical potential  $\mu_5$ , since systems produced in high-energy collisions are often in a non-equilibrium state \cite{rr32,rr33,rr34}, and the presence of chiral asymmetry further enhances the complexity of the system. Thus, integrating Tsallis statistics into the NJL model makes it possible to study the QCD phase structure in complex environments, including those with chiral imbalance.
	
	In this work, we use the two-flavor NJL model to investigate the chiral phase transition temperature as a function of the magnetic field $eB$, the chiral chemical potential $\mu_5$, and the non-extensive parameter $q$. The paper is structured as follows: In Section \ref{sec:02 setup}, we introduce the two-flavor NJL model with chiral imbalance and strong magnetic field, and derive the thermodynamic potential using Tsallis statistics. Section \ref{sec:03 setup} presents numerical results on the phase transition temperature and thermodynamic quantities under various values of $q$, $eB$, and $\mu_5$. We conclude with a summary and outlook in Section \ref{sec:04 summary}.
	
	\section{Theoretical Model}\label{sec:02 setup}
	The Lagrangian density for the NJL model incorporating both a strong magnetic field and finite chiral chemical potential \cite{rr54,rr55,rr56} is given by
	
	\begin{equation}
		\begin{array}{c}{L=\overline{\psi}\left(i\gamma_{\mu}D^{\mu}+\mu\gamma^{0}+\mu_{5}\gamma^{0}\gamma^{5}\right)\psi+
G[(\overline{\psi}\psi)^{2}+(\overline{\psi}i\gamma^{5}\tau\psi)^{2}]}\end{array} ,
	\end{equation}
	where $G$ represents the coupling constant, $\mu$ denotes the quark chemical potential, and the covariant derivative embeds the quark coupling to the external magnetic field. To investigate this system, we employ Tsallis statistics as a generalization of Boltzmann--Gibbs (B-G) statistics. This framework enables the study of the thermodynamic potential \cite{rr57} in the SU(2)-NJL model with chiral chemical potential $\mu_5$ , characterized by the non-extensive real parameter $q$. $\mu_5$ denotes the quark chiral chemical potential, which can effectively model the topological fluctuations of chiral charge $N_{5}=N_{\mathrm{R}}-N_{\mathrm{L}}$ induced by QCD axial anomaly, as theoretically predicted.
	
	The exponential form defining the parameter $q$ \cite{rr45,rr46,rr58} is given by
	\begin{equation}
		\exp_q\left(x\right)\equiv\begin{cases}\left(1+\left(q-1\right)x\right)^{1/(q-1)}  ~~x>0, \\\left(1+\left(1-q\right)x\right)^{1/(1-q)} ~~x\leq0, &\end{cases}
	\end{equation}
    where $x=\frac{E-\mu}{T}$ is defined, and in the limit $q\to1$ , the standard exponential form $\lim_{q\to1}\exp_{q}(x)=\exp(x)$ is recovered. The inverse function of the  $q$-logarithm is expressed as:
    \begin{equation}
    	\ln_q(x)\equiv\begin{cases}\frac{x^{q-1}-1}{q-1} ~~x>0, \\\frac{x^{1-q}-1}{1-q} ~~x\leq0. \end{cases}
    \end{equation}

    The distribution function for fermions at temperature $T$ and chemical potential $\mu$ is given as
    \begin{equation}\label{eq4}
    	f_{\pm}(E,\mu)=\begin{cases}\frac{1}{1+(1+\frac{(q-1)(E\mp\mu)}{T})^{\frac{1}{q-1}}} ~~x>0, \\\frac{1}{1+(1+\frac{(1-q)(E\mp\mu)}{T})^{\frac{1}{1-q}}} ~~x\leq0. \end{cases}
    \end{equation}

    The thermodynamic potential is obtained as
    \begin{equation}\label{eq5}
    	\begin{aligned}\Omega\left(T,\mu_{5},B,q,\mu\right)&=\frac{\sigma^2}{4G}-N_c\sum_{f=u,d}\frac{\left|q_fB\right|}{2\pi}\sum_{s,n}\alpha_{sn}
    \int_{-\infty}^\infty\frac{dp_z}{2\pi}E_{f,n,s}\left(\mathbf{p}\right)-N_cT\sum_{f=u,d}\frac{\left|q_fB\right|}{2\pi}\\&\times\sum_{s,n}\alpha_{sn}\int_{-\infty}^\infty
    \frac{dp_z}{2\pi}\begin{cases}\frac{[1+(1-\varphi^{+})^{\frac{1}{1-q}}]^{q-1}-1}{q-1}+\frac{[1+(1-\varphi^{-})^{\frac{1}{1-q}}]^{q-1}-1}{q-1} ~x>0,
    \\\\\frac{[1+(1+\varphi^{+})^{\frac{1}{q-1}}]^{q-1}-1}{q-1}+\frac{[1+(1+\varphi^{-})^{\frac{1}{q-1}}]^{q-1}-1}{q-1} ~x\leq0,&\end{cases}\end{aligned}
    \end{equation}
    where
    \begin{equation}
    	\varphi^{\pm}=\frac{(q-1)(E_{f,n,s}\left(\mathbf{p}\right) \mp\mu)}{-T},
    \end{equation}
    and $\sigma$ the chiral condensate, and the dynamical quark mass $M = \sigma$. The specific form of the energy expression is $E_{f,n,s}^2=M_f^2+(\left|\mathbf{p}\right|+s\mu_5\mathrm{sgn}\left(p_z\right))^2$, $\mathbf{p}^2 = p_z^2+2\begin{vmatrix}{q}_fB\end{vmatrix}{n}$, ~$s=\pm1$, ~where $n=0,1,2...$ represents the Landau energy level. \(N_{c}=3\) is the number of colors, $q_u = \frac{2}{3}e$ and $ q_d=-\frac{1}{3}e$ are the charge of $u$ and $d$ quarks, respectively. The spin degeneracy factor takes the values as
    \begin{equation}
    	\alpha_{sn}=\begin{cases}\delta_{s,+1}, \enspace for\enspace n = 0,\enspace q_fB>0,\\\delta_{s,-1},\enspace for\enspace n=0,\enspace q_fB<0,\\\enspace 1,\quad\;\, for\enspace n\neq0.\end{cases}
    \end{equation}

    Note that the vacuum part of \(\Omega\left(T,\mu_{5},B,q,\mu\right)\) exhibits ultraviolet divergence and requires regularization. The soft-cutoff scheme \cite{rr55} is adopted in this work. The soft-cutoff is implemented by using the following formula as
    \begin{equation}
    	f_\Lambda\left(\mathbf{p}\right)=\sqrt{\frac{\Lambda^{2N}}{\Lambda^{2N}+\mid \mathbf{p}\mid^{2N}}}.
    \end{equation}

    By taking $N=5$, one can obtain the thermodynamic potential of the vacuum part as
    \begin{equation}
    \Omega_{0}(\mu_5,B)=-N_c\sum_{f=u,d}\frac{\left|q_fB\right|}{2\pi}\sum_{s,n}\alpha_{sn}\int_{-\infty}^\infty
    \frac{dp_z}{2\pi}f_\Lambda^2(\mathbf{p})E_{f,n,s}(\mathbf{p}).
    \end{equation}

    By minimizing the thermodynamic potential, one can obtain the gap equation as
    \begin{equation}
    	\frac{\partial\Omega\left(T,\mu_{5},B,q,\mu\right)}{\partial M}=0,\quad\frac{\partial^2\Omega\left(T,\mu_{5},B,q,\mu\right)}{\partial M^2}>0. 
    \end{equation}

    At the same time, we also investigated another regularization scheme, the medium separation scheme (MSS) from literature \cite{rr59} was applied to process them, and the gap equation \(M = m-2G\sum_f\left\langle\bar{\psi}_f\psi_f\right\rangle\)
    was used to solve for the quark mass. In the literature, the expression for the quark condensate is

    \begin{equation}
    	\begin{gathered}\left\langle\bar{\psi}_{f}\psi_{f}\right\rangle_{B\neq0}^{\mu_{5}\neq0}=-\frac{N_c\left|q_f\right|B}{\left(2\pi\right)^2}\int dp_z \frac{M_f}{E_{f,0,s}\left(p\right)}\Big[1-f_{FD}\Big(E_{f,0,s}\left(p\right)-\mu\Big)-f_{FD}\Big(E_{f,0,s}\left(p\right)+\mu\Big)\Big]\\+\sum_{n=1}^{\infty}\sum_{s}\int dp_{z} \frac{M_{f}}{E_{f,n,s}\left(p\right)}\Big[1-f_{FD}\Big(E_{f,n,s}\left(p\right)-\mu\Big)-f_{FD}\Big(E_{f,n,s}\left(p\right)+\mu\Big)\Big] \end{gathered}.
    \end{equation}

    This can be simplified and, by incorporating parameter $q$ using Eq.~\ref{eq4}, it transforms into

    \begin{equation}\label{eq12}
    	\begin{aligned}\left\langle\overline{\psi}_f\psi_f\right\rangle_{B\neq0}^{\mu_5\neq0}&=-\frac{M_f\left(M_{0_f}^2-M_f^2+2\mu_5^2\right)}{2}I_{1_{\log}}+I_{1_{\mathrm{finitel}}}+I_{1_{\mathrm{finite}2}}+I_{2_{\mathrm{finite}}}+I_{\mathrm{quad}}
    	\\&+\frac{N_c\left|q_f\right|B}{2\pi^2}\Bigg[\int_{-\infty}^\infty dp_z\frac{M_f}{E_{f,0,s}\left(p\right)}\begin{cases}\frac{1}{1+(1+\frac{\left(q-1\right)\left(E_{f,0,s}\mp\mu\right)}{T})^{\frac{1}{q-1}}},&x>0\\\frac{1}{1+(1+\frac{\left(1-q\right)\left(E_{f,0,s}\mp\mu\right)}{T})^{\frac{1}{1-q}}},&x\leq0\end{cases}
    	\\&+\sum_{n=1}^\infty\sum_s\int_{-\infty}^\infty dp_z\frac{M_f}{E_{f,0,s}\left(p\right)}\begin{cases}\frac{1}{1+(1+\frac{\left(q-1\right)\left(E_{f,n,s}\mp\mu\right)}{T})^{\frac{1}{q-1}}},&x>0\\\frac{1}{1+(1+\frac{\left(1-q\right)\left(E_{f,n,s}\mp\mu\right)}{T})^{\frac{1}{1-q}}},&x\leq0\end{cases}\Bigg]
    	\end{aligned},
    \end{equation}
    where
    \begin{equation}
    	I_{1_{\log}}=\frac{N_c\left|q_f\right|B}{\left(2\pi\right)^2}\sum_{n=0}^\infty\sum_s\int dp_z\frac{1}{\varepsilon_{f,n,s}^3},
    \end{equation}
    \begin{equation}
    	I_{1_{\mathrm{finitel}}}=-\frac{N_c\left|q_f\right|B}{\left(2\pi\right)^2}\sum_{n=0}^\infty\sum_s\int dp_z\left(\frac{3}{8}\right)\frac{(M_fA^2-4M_f\mu_s^2M_{0_f}^2)}{\varepsilon_{f,n,s}^5},
    \end{equation}
    \begin{equation}
    	I_{1_{\mathrm{finte2}}}=-\frac{N_c\left|q_f\right|B}{\left(2\pi\right)^2}\left(\frac{15}{16}\right)\sum_{n=0}^\infty\sum_s\int dp_z
    	\int_0^1dx\frac{\left(1-x\right)^2M_f\left(A+2s\mu_5\sqrt{p_z^2+2n\left|q_f\right|B}\right)^3}{\left[\varepsilon_{f,n,s}^2-x\left(A+2s\mu_5
    \sqrt{p_z^2+2n\left|q_f\right|B}\right)\right]^{7/2}},
    \end{equation}
    \begin{equation}
    	I_{2_{\mathrm{finite}}}=\left(\frac{1}{2}\right)\frac{N_c\left|q_f\right|B}{\left(2\pi\right)^2}\int dp_z\int_0^1dx\frac{M_f\left(A+2p_z\mu_5\right)}{\left[\varepsilon_{f,0,s}^2-x(A+2p_z\mu_5)\right]^{3/2}},
    \end{equation}
    \begin{equation}
    	I_{\mathrm{quad}}=\frac{M_f}{M_{0_f}}\left[-\frac{N_c\left|q_f\right|B}{\left(2\pi\right)^2}\sum_{n=0}^\infty\sum_s\int dp_z\frac{M_{0_f}}{\varepsilon_{f,n,s}}+\frac{N_c\left|q_f\right|B}{\left(2\pi\right)^2}\int dp_z\frac{M_{0_f}}{\varepsilon_{f,0,s}}\right],
    \end{equation}
    with $A=M_{0_f}^2-M_f^2-\mu_5^2$, $E_{f,0,s}^2=M_f^2+\left(p_z-\mu_5\right)^2$, $\varepsilon_{f,n,s}=\sqrt{M_{0_f}^2+p_z^2+2n\left|q_f\right|B}$, $\varepsilon_{f,0,s}=\sqrt{M_{0_f}^2+p_z^2}$.

    Let's analyze the corresponding thermodynamic quantity, such as pressure, which is equal to the negative of the thermodynamic potential, i.e. $P = -\Omega$. By subtracting vacuum pressure ( $P_0 = -\Omega_0 $ ) contribution from the total pressure, one utilizes the normalized pressure obtained as \cite{rr31,rr54}
    \begin{equation}
    	P_{\mathrm{eff}}\left(T,\mu_5,B,q,\mu\right)=-\Omega\left(T,\mu_5,B,q,\mu\right)+\Omega_0\left(\mu_5,B\right),
    \end{equation}
    and the transverse pressure and longitudinal pressure \cite{rr60} are given as
    \begin{equation}
    	P_{\parallel}\left(T,\mu_{5},B,q,\mu\right)=P_{eff}\left(T,\mu_{5},B,q,\mu\right),
    \end{equation}
    \begin{equation}
    	P_\perp\left(T,\mu_5,B,q,\mu\right)=P_\parallel\left(T,\mu_5,B,q,\mu\right)-\mathcal{M}B,
    \end{equation}
    where $ \mathcal{M} $ represents the magnetization as
    \begin{equation}
    	\mathcal{M}=\frac{\partial P_{eff}\left(T,\mu_5,B,q,\mu\right)}{\partial B}.
    \end{equation}

    The speed of sound squared $c_s^2$ taken as an essential fundamental quantity that is employed in discussions for hot QCD medium, so we have
    \begin{equation}
    	c_s^2=\frac{\partial P}{\partial\varepsilon}{\Big|}_S=\frac{\partial P_{eff}\left(T,\mu_5,B,q,\mu\right)}{\partial T}{\Big|}_{V}{\Big /} \left ( T\frac{\partial^2P_{eff}\left(T,\mu_5,B,q,\mu\right)}{\partial T^2} \right ){\Big|}_{V}, 
    \end{equation}
    where \(P\) and \(\varepsilon\) are the total pressure and energy density, summing over all charged particles (\(u\) and \(d\) quarks).

	\section{NUMERICAL RESULTS}\label{sec:03 setup}
	To better study the phase diagram structure of QCD under the Tsallis statistical model, we selected a system with a quark chemical potential $\mu = 0$. By fitting the meson decay constant $f_{\pi}=92.4\;~\textrm{MeV}$ and the vacuum chiral condensate $\left\langle\overline{u}u\right\rangle^{\frac{1}{3}} = -245.7~\textrm{MeV}$, we adopt the parameters  $m_{u}=m_{d}=6~\textrm{MeV}$, $\Lambda = 620~\textrm{MeV},\;G\Lambda^2=2.2$ in the SU(2) magnetized system.
	
	\subsection{The phase diagram structure of QCD under the Tsallis statistical model}	
\begin{figure}[tp]
		\centering
		\includegraphics[width=1.0\linewidth]{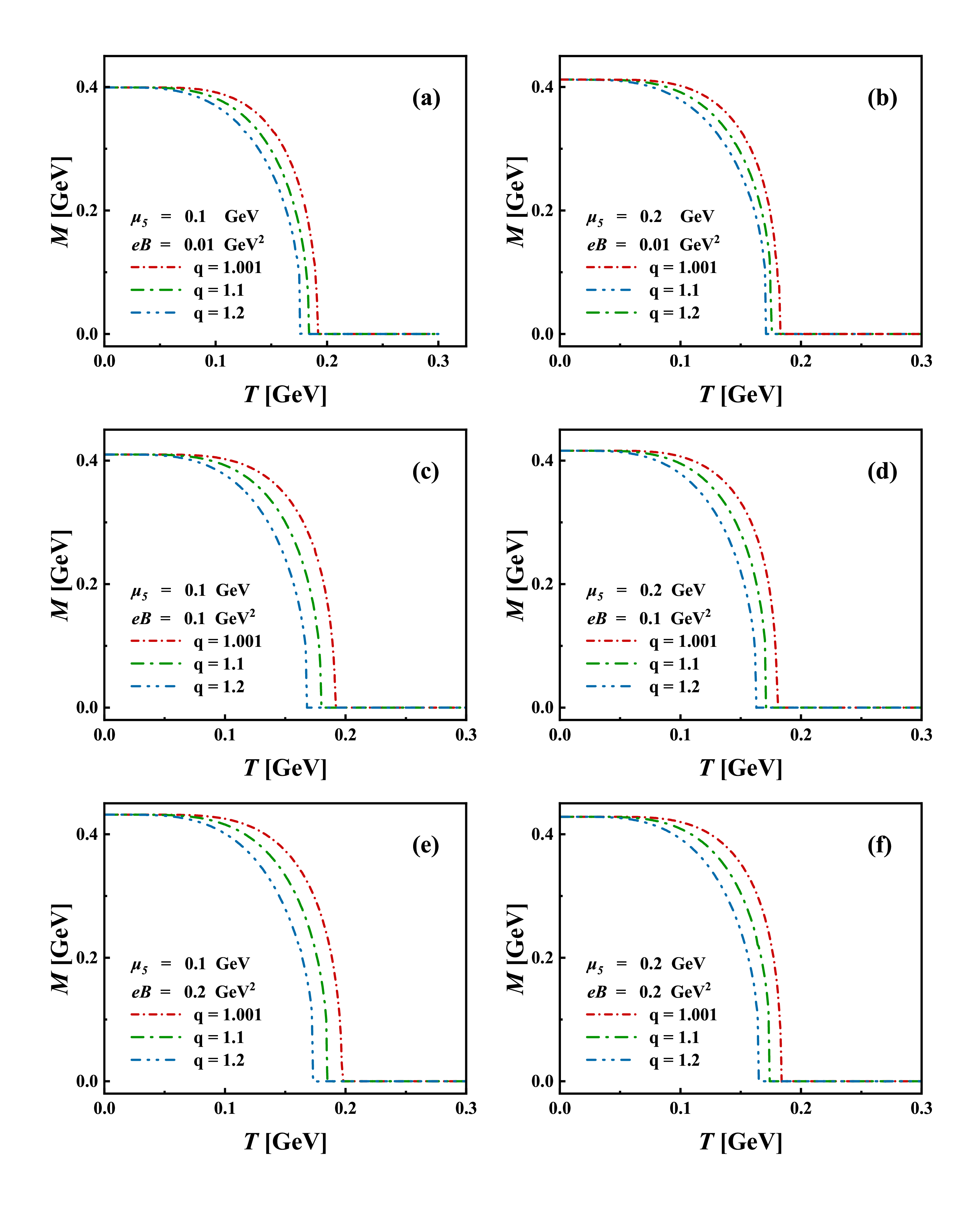}
		\caption{The dependence of the quark dynamical mass \(M\) on temperature \(T\) for different Tsallis parameter \(q\) values. For magnetic field \(eB=0.01 ~\textrm{GeV}^2\): (a) \(\mu_{5}=0.1~\textrm{GeV}\), (b) \(\mu_{5}=0.2~\textrm{GeV}\); for magnetic field \(eB=0.1 ~\textrm{GeV}^2\): (c) \(\mu_{5}=0.1~\textrm{GeV}\), (d) \(\mu_{5}=0.2~\textrm{GeV}\); for magnetic field \(eB=0.2 ~\textrm{GeV}^2\): (e) \(\mu_{5}=0.1~\textrm{GeV}\), (f) \(\mu_{5}=0.2~\textrm{GeV}\).}
		\label{fig:1}
	\end{figure}

	Using the dynamical quark mass obtained from the gap equation, we systematically analyze the influence of various parameters on the dynamical quark mass during the chiral phase transition. The chiral chemical potential \(\mu_5\) is varied in the range of \(0.1 \sim 0.2\ \text{GeV}\), the magnetic field strength \(eB\) is varied in the range of \(0.01 \sim 0.2\ \text{GeV}^2\), and the non-extensive parameter \(q\) takes the values \(1.001\), \(1.1\), and \(1.2\), respectively. Figure \ref{fig:1} shows the temperature dependence of the dynamical quark mass. It is evident from Fig.~\ref{fig:1} that the parameter \(q\) significantly modifies the temperature at which the dynamical quark mass drops sharply, i.e., the critical temperature \(T_c\), reducing it by approximately \(0.01\ \text{GeV}\). This abrupt change in the quark mass is a characteristic of a second-order phase transition.
	
\begin{figure}[tp]
		\centering
		\includegraphics[width=1.0\linewidth]{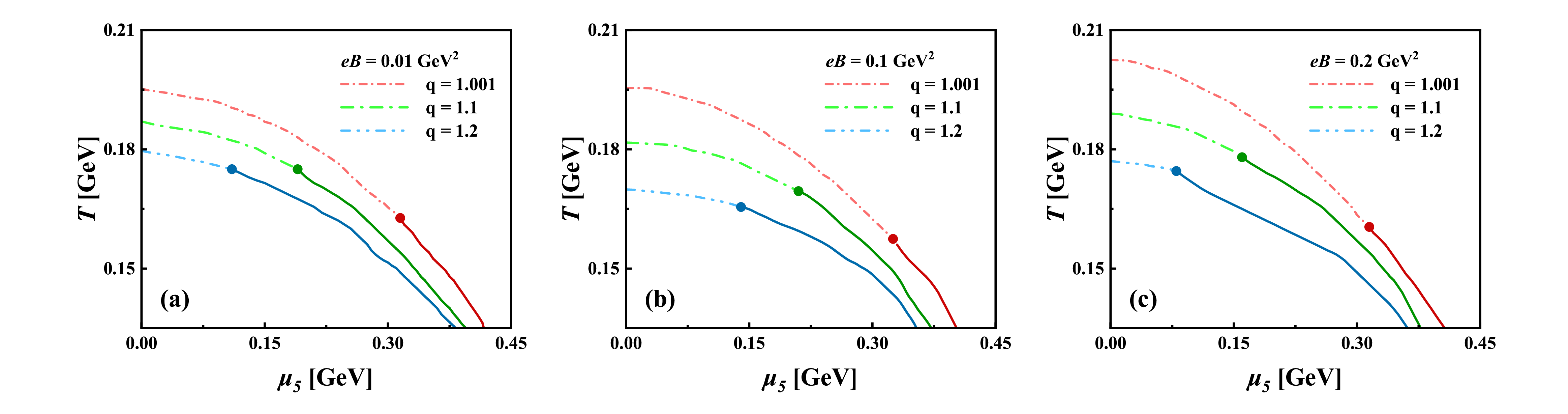}
		\caption{Dependence of the critical temperature \(T_c\) on the chiral chemical potential \(\mu_5\)
		under different Tsallis parameters \(q\), where (a) \(eB=0.01 ~\textrm{GeV}^2\); (b) \(eB=0.1 ~\textrm{GeV}^2\); (c) \(eB=0.2 ~\textrm{GeV}^2\).}
		\label{fig:2}
	\end{figure}

\begin{figure}[tp]
		\centering
		\includegraphics[width=1.0\linewidth]{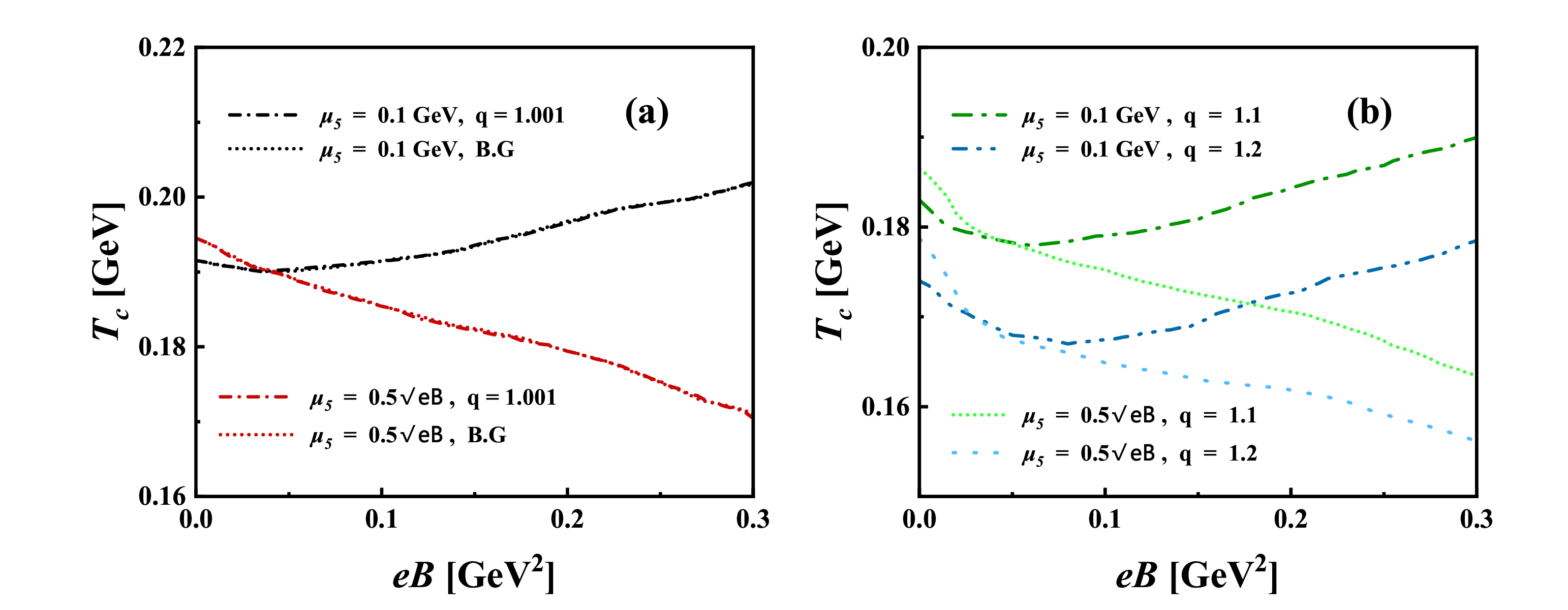}
		\caption{Dependence of the critical temperature \(T_c\) on the magnetic field $eB$
			under different Tsallis parameters $q$: (a) Comparison between Boltzmann-Gibbs statistics
			(B.G.) and the case $q = 1.001 $, with fixed $\mu_{5} = 0.1~\textrm{GeV}$ and field-dependent
			$\mu_{5}=0.5\sqrt{eB}$; (b) Behavior under $q = 1.1$ and $ 1.2 $ for the same two choices of $\mu_5 $.}
		\label{fig:3}
	\end{figure}

\begin{figure}[tp]
	    \centering
	   	\includegraphics[width=0.5\linewidth]{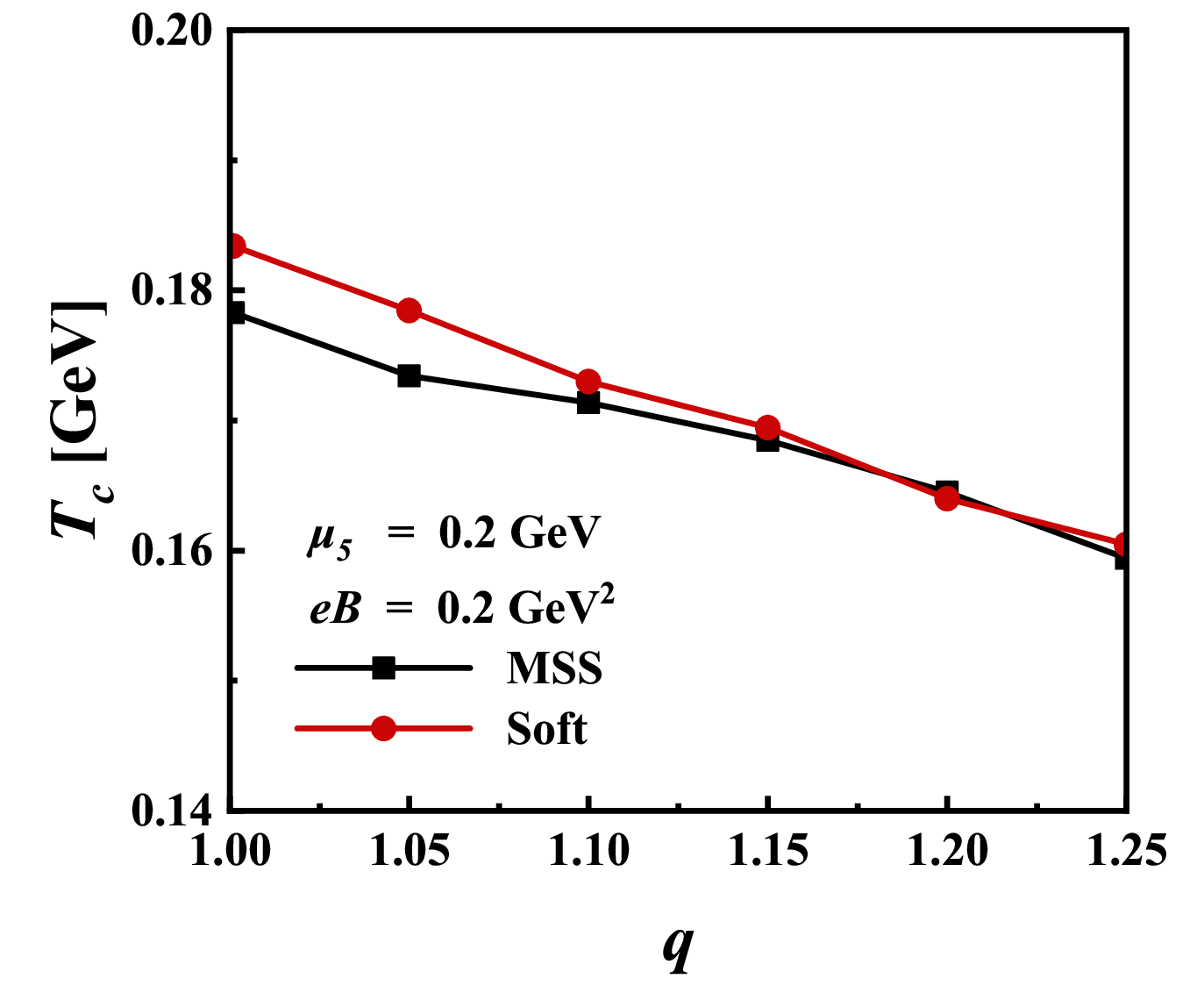}
	   	\caption{The dependence of the critical temperature \(T_c\) on the Tsallis parameter \( q \) is shown for \( \mu_5 = 0.2 \, \text{GeV} \) and \( eB = 0.2 \, \text{GeV}^2 \), comparing results obtained using the soft-cutoff and MSS schemes.}
	   	\label{fig:4}
    \end{figure}

	By comparing panels (a) with (b), (c) with (d), and (e) with (f), it is found that the chiral chemical potential \(\mu_5\) has a notable effect on the chiral phase transition temperature, which decreases with increasing \(\mu_5\). In contrast, comparisons between panels (a) and (c), as well as (c) and (e), reveal that the magnetic field exhibits an opposite trend compared to \(\mu_5\), namely the transition temperature slowly increases with increasing magnetic field. Furthermore, an increase in the magnetic field strength leads to a more significant enhancement of the initial quark mass compared to the case of \(\mu_5\).To isolate the individual effects of each variable on the chiral phase transition temperature, we further analyze the roles of \(\mu_5\), \(eB\), and \(q\) on \(T_c\) in Figs.~\ref{fig:2}, \ref{fig:3}, and \ref{fig:4}, respectively.
	
In Fig.~\ref{fig:2}, the phase diagram of the chiral critical temperature \(T_c\) as a function of the chiral chemical potential \(\mu_5\) is presented. It can be observed that, regardless of the external magnetic field strength or the value of the non-extensive parameter \(q\), \(T_c\) decreases monotonically and smoothly with increasing \(\mu_5\), exhibiting a behavior similar to inverse magnetic catalysis (IMC). Taking a fixed magnetic field \(eB = 0.1\ \text{GeV}^2\) (as shown in Fig.~\ref{fig:2}(b)) as an example, as \(\mu_5\) increases from its initial value, the reduction in the phase transition temperature already reaches about $0.01$ GeV in the initial decreasing stage. Furthermore, for a fixed \(\mu_5\), as the parameter q increases, \(T_c\) also exhibits a decreasing trend, again reflecting IMC characteristics. Meanwhile, the value of \(\mu_5\) corresponding to the critical end point (CEP) also shifts significantly toward lower chemical potential with increasing q, indicating that the parameter \(q\) plays a significant role in modulating the location of the CEP associated with the phase transition abrupt point.

A horizontal comparison across Fig.~\ref{fig:2}(a)$\rightarrow$(c) shows that the magnetic field strength has a weak influence on the overall trend of the \(T_c\) - \(\mu_5\) curve. When \(q = 1.001\) and the external magnetic field increases, \(T_c\) exhibits a slight rise, displaying the feature of magnetic catalysis (MC), indicating that the system is in the MC-dominated region rather than the IMC region. This demonstrates that although an increase in \(\mu_5\) suppresses \(T_c\) (inducing IMC), the magnetic field can partially counteract this suppression. This phenomenon can be attributed to the fact that the power-law statistical distribution (non-extensive statistics) weakens the shielding effect of thermal fluctuations on the magnetic field, thereby maintaining the enhancing effect of the lowest Landau level (LLL) on the quark condensate. For a fixed magnetic field strength, \(\mu_5\) introduces chiral charge imbalance and an effective chemical potential shift, which counteract the catalytic effect of the magnetic field and drive the system into the IMC phase.

A further comparison of the behavior of the critical end point (CEP) reveals that for \(q = 1.001\), the value of \(\mu_5\) corresponding to the CEP remains almost unchanged with the magnetic field, stabilizing at about \(0.315\ \text{GeV}\). For \(q = 1.1\) and \(q = 1.2\), the influence of the magnetic field on the CEP location becomes significant, markedly altering the value of \(\mu_5\) corresponding to the CEP. For example, when \(q = 1.1\), as the magnetic field varies, the values of \(\mu_5\) corresponding to the CEP are approximately \(0.19\ \text{GeV}\), \(0.21\ \text{GeV}\), and \(0.16\ \text{GeV}\), respectively, exhibiting a non-monotonic evolution that first increases and then decreases.

The phenomenon that the critical \(\mu_5\) value at the CEP remains nearly unchanged under different magnetic fields for \(q = 1.001\) is 
explained in Fig.~\ref{fig:3}(a). In that figure, we compare the results obtained from B.G. statistics with those from \(q = 1.001\). The two sets of results are numerically identical, verifying the consistency between the two statistical approaches in treating the NJL model, as reported in Refs.~\cite{rr47,rr57}. When \(q \to 1\), the Tsallis entropy recovers extensivity and naturally reduces to the standard Boltzmann-Gibbs statistical form.

	Fig.~\ref{fig:3} highlights the variation of the critical temperature \(T_c\) with the magnetic field $eB$  under two different chiral chemical potential conditions: a fixed $\mu_{5} = 0.1~\textrm{GeV}$ and a magnetic-field-dependent form $\mu_{5}=0.5\sqrt{eB}$. In the near-equilibrium scenario ($q\approx1 $), the results from Tsallis statistics resemble those of conventional Fermi statistics ~\cite{rr57}: as shown in Fig.~\ref{fig:3}(a), for fixed $\mu_{5} = 0.1~\textrm{GeV}$ , $T_{c} $ increases with $eB$, exhibiting an effect analogous to magnetic catalysis (MC). In contrast, when $\mu_5$  takes the magnetic-field-dependent form $\mu_{5}=0.5\sqrt{eB}$ , $T_{c} $ decreases with increasing $eB$, displaying characteristics analogous to inverse magnetic catalysis (IMC). Under non-equilibrium conditions ($q > 1 $): Fig.~\ref{fig:3}(b) reveals non-monotonic behavior in the weak magnetic field region. For the case with fixed $\mu_{5} = 0.1~\textrm{GeV}$, $T_{c} $ first decreases with $eB$ and then rises as the magnetic field strengthens further. For the case with $\mu_{5}=0.5\sqrt{eB}$, \(T_c\) decreases monotonically with $eB$, and the decline is more pronounced in weaker magnetic fields.

	
	Based on the results shown in Fig.~\ref{fig:2} and \ref{fig:3}, we further analyze the influence of the non-extensive parameter \( q \) on theoretical predictions. The study shows that a larger \( q \) significantly enhances the non-monotonic features in the transverse momentum spectrum and amplifies the non-equilibrium response of the system in a magnetic field environment. This phenomenon directly reflects the intrinsic non-extensive nature of the Tsallis statistical framework.  We conducted a detailed analysis using the medium separation scheme (MSS) scheme given by Eq.~\ref{eq12}, a comparison between the two cutoff schemes (MSS and soft-cutoff) is presented in Fig.~\ref{fig:3}. With fixed magnetic field strength \( eB = 0.2 \, \text{GeV}^2 \) and chiral chemical potential \( \mu_5 = 0.2 \, \text{GeV} \), the critical temperature \(T_c\) shows a monotonic decreasing trend as the non-extensive parameter \( q \) increases. Notably, for \( q \) values between 1.1 and 1.2, the results from both schemes are in excellent agreement, with negligible differences. This convergence indicates that within the non-extensive statistical framework, the parameter \( q \) plays a dominant role in determining the system’s behavior, while the choice of cutoff scheme has minimal impact within this parameter range. In summary, although different cutoff methods differ fundamentally in their physical mechanisms and lead to variations in data trends, the physical picture they reveal still possesses a certain intrinsic consistency, reflecting the response characteristics of the QCD phase structure under non-extensive statistics to different microscopic treatments. Therefore, to systematically investigate the regulatory effect of the non-extensive parameter \( q \) on theoretical outcomes, in the second part of this paper, concerning the calculation of thermodynamic potentials and related physical properties, we continue to primarily employ the soft truncation scheme for subsequent analysis.
	
	\subsection{The thermodynamic properties under the Tsallis statistical model}
	
	The chiral particle number density \cite{rr13}\cite{rr61,rr62,rr63} can be defined as
	\begin{equation}
		n_5=-\frac{\partial \Omega}{\partial\mu_5}.
	\end{equation}
	
\begin{figure}[tp]
		\centering
		\includegraphics[width=1.0\linewidth]{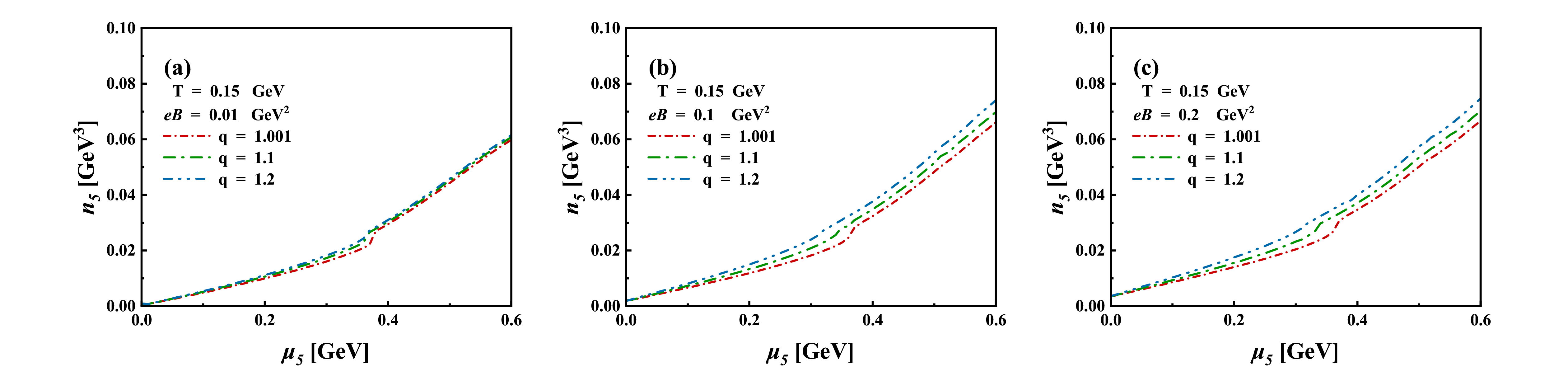}
		\caption{Chiral number density \(n_5\) as a function of the chiral chemical potential \(\mu_5\) at fixed temperature \(T = 0.15\ \text{GeV}\) for different Tsallis parameters \(q\) and different magnetic field conditions, where (a) \(eB = 0.01\ \text{GeV}^2\), (b) \(eB = 0.1\ \text{GeV}^2\), and (c) \(eB = 0.2\ \text{GeV}^2\).}
		\label{fig:5}
	\end{figure}

As shown in Fig.~\ref{fig:5}, at a fixed temperature \(T = 0.15\ \text{GeV}\), under different magnetic field values, the chiral number density \(n_5\) increases with the chiral chemical potential \(\mu_5\), indicating that a higher chiral chemical potential enhances the chiral imbalance of the system. Under low magnetic field (\(eB = 0.01\ \text{GeV}^2\)) conditions, the growth trend of \(n_5\) with \(\mu_5\) is less affected by the Tsallis parameter \(q\); as the magnetic field strength increases, the influence of the parameter \(q\) gradually becomes significant. The first-order phase transition feature is most evident at \(q = 1.001\), and the presence of the magnetic field further enhances this transition feature. Furthermore, the larger the value of \(q\), the faster \(n_5\) increases with \(\mu_5\), implying that stronger non-equilibrium conditions promote a more rapid response of the chiral number density to the chiral chemical potential.
	
\begin{figure}[tp]
		\centering
		\includegraphics[width=1.0\linewidth]{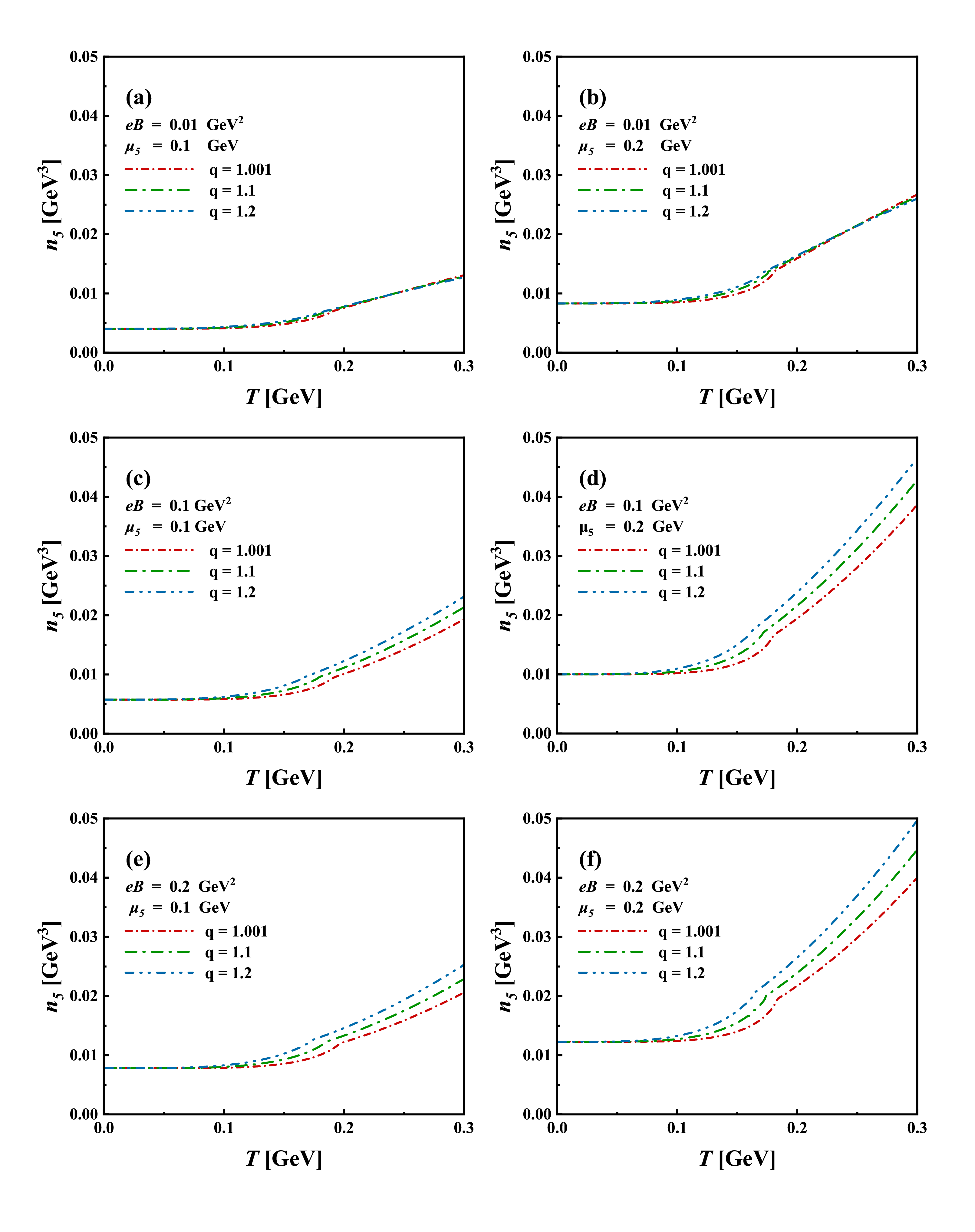}
		\caption{Chiral number density \(n_5\) as a function of temperature \(T\) for different Tsallis parameters \(q\). For magnetic field \(eB=0.01 ~\textrm{GeV}^2\): (a) \(\mu_{5}=0.1~\textrm{GeV}\), (b) \(\mu_{5}=0.2~\textrm{GeV}\); for magnetic field \(eB=0.1 ~\textrm{GeV}^2\): (c) \(\mu_{5}=0.1~\textrm{GeV}\), (d) \(\mu_{5}=0.2~\textrm{GeV}\); for magnetic field \(eB=0.2 ~\textrm{GeV}^2\): (e) \(\mu_{5}=0.1~\textrm{GeV}\), (f) \(\mu_{5}=0.2~\textrm{GeV}\).}
		\label{fig:6}
	\end{figure}

Figure \ref{fig:6} shows, under non-extensive Tsallis statistics, the temperature dependence of the chiral number density \(n_5\), where the chiral chemical potential takes two values (\(\mu_{5}=0.1~\textrm{GeV}\) and \(0.2~\textrm{GeV}\)), and the effects of different \(q\) parameters and different magnetic field values are considered. The results show that, for the two different \(\mu_5\) values, in the low-temperature region (\(T < 0.1~\textrm{GeV}\)), \(n_5\) remains nearly constant and very small, forming an obvious plateau; as \(\mu_5\) increases, this plateau region gradually narrows. Subsequently, \(n_5\) increases monotonically with temperature. A small jump appears near the critical temperature \(T_c\), indicating that the system exhibits a first-order phase transition feature. This transition feature is most evident at \(q = 1.001\), and as the value of \(q\) increases, the first-order transition feature gradually becomes smooth. Furthermore, a larger Tsallis parameter \(q\) enhances the monotonic increasing trend of \(n_5\) with \(T\). A higher \(\mu_{5}=0.2~\textrm{GeV}\) significantly increases the magnitude of \(n_5\) over the entire temperature range.

Further comparing the results of subplots (a), (c), and (e) in Fig.\ref{fig:6}, it can be found that under low magnetic field (\(eB = 0.01\ \text{GeV}^2\)) conditions, the growth trend of \(n_5\) with \(T\) is less affected by the Tsallis parameter \(q\), with its main effect being only a slight increase in the initial value of the constant \(n_5\) region. In summary, it can be concluded that \(\mu_5\) is the dominant factor affecting the chiral number density, the magnetic field effect is negligible, and the parameter \(q\) has the characteristic of suppressing the first-order phase transition feature.

	\begin{figure}[tp]
		\centering
		\includegraphics[width=1.0\linewidth]{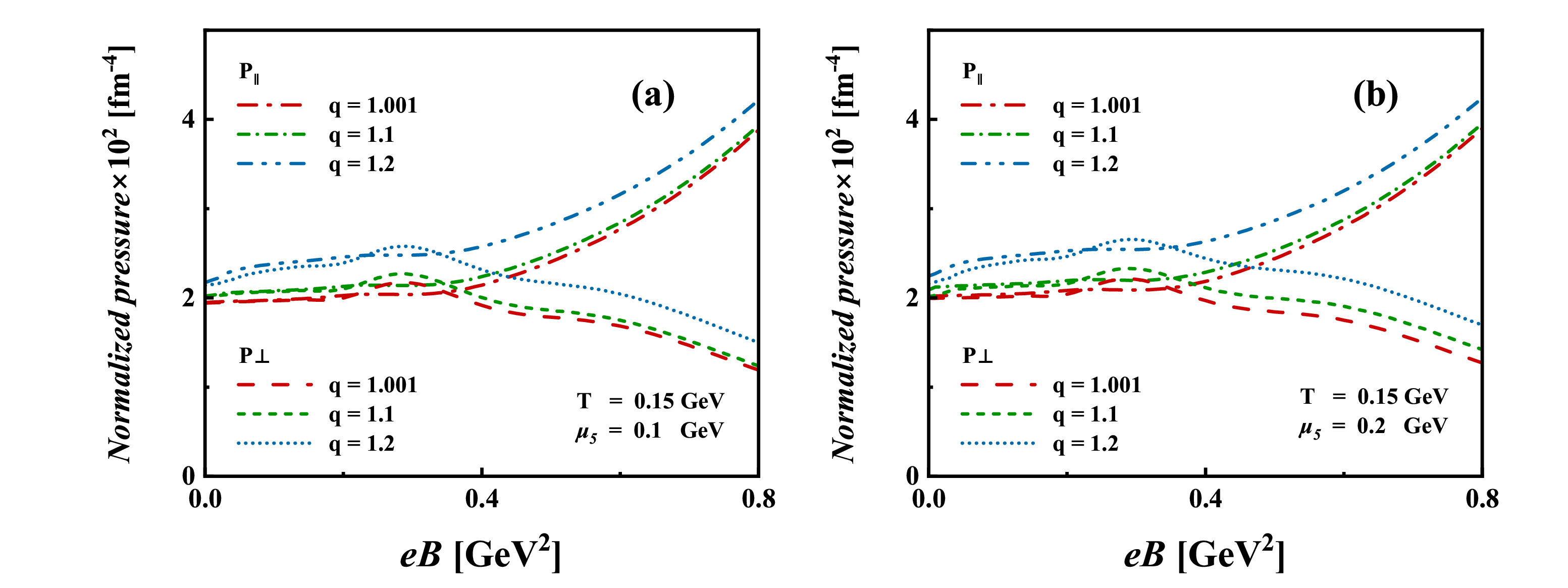}
		\caption{Pressure anisotropy under strong magnetic fields at fixed temperature $T=0.15~\textrm{GeV}$. Shown are the longitudinal pressure $P_{\parallel}$ (parallel to $eB$) and transverse pressure $P_{\perp}$ (perpendicular to $eB$), for different chiral chemical potentials: (a) $\mu_{5}=0.1~\textrm{GeV}$, (b) $\mu_{5}=0.2~\textrm{GeV}$.}
		\label{fig:7}
	\end{figure}
	\begin{figure}[tp]
		\centering
		\includegraphics[width=1.0\linewidth]{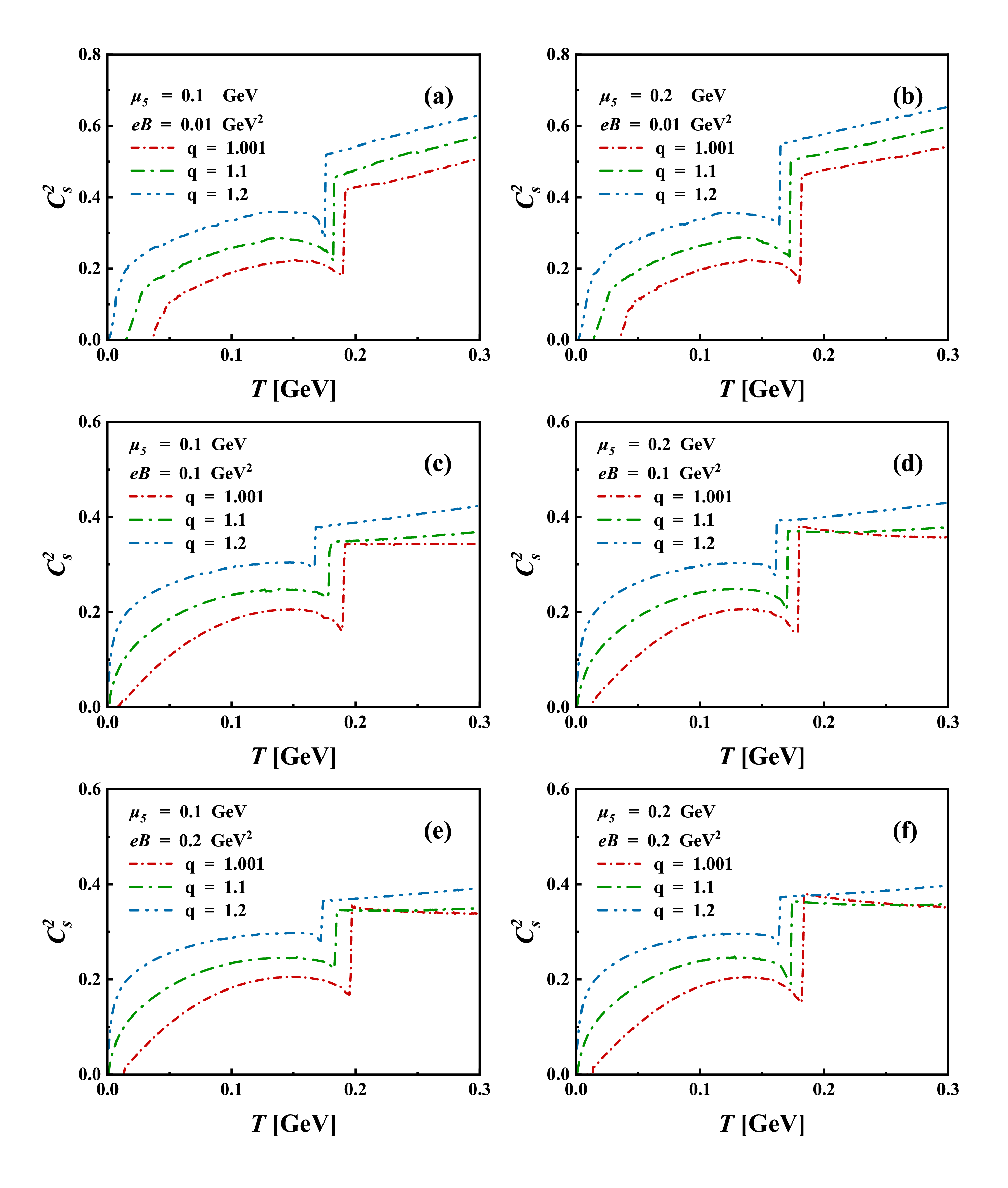}
		\caption{The squared speed of sound $c_s^2$ as a function of temperature $T$ for different values of the Tsallis parameter $q$. For magnetic field \(eB=0.01 ~\textrm{GeV}^2\): (a) \(\mu_{5}=0.1~\textrm{GeV}\), (b) \(\mu_{5}=0.2~\textrm{GeV}\); for magnetic field \(eB=0.1 ~\textrm{GeV}^2\): (c) \(\mu_{5}=0.1~\textrm{GeV}\), (d) \(\mu_{5}=0.2~\textrm{GeV}\); for magnetic field \(eB=0.2 ~\textrm{GeV}^2\): (e) \(\mu_{5}=0.1~\textrm{GeV}\), (f) \(\mu_{5}=0.2~\textrm{GeV}\).}
		\label{fig:8}
	\end{figure}
	
Figure \ref{fig:7} shows, within the non-extensive Tsallis statistical framework, the anisotropic pressure response under a strong magnetic field at temperature \(T = 0.15\ \text{GeV}\). The results indicate that the influence of \(\mu_5\) on the normalized pressure is negligible. The longitudinal pressure \(P_{\parallel}\) increases monotonically with the magnetic field strength \(eB\), while the transverse pressure \(P_{\perp}\) exhibits non-monotonic behavior: it gradually rises in the weak-field region and then turns to decrease in the strong-field region. This anisotropy originates from the competition mechanism between Landau quantization (dominant in the low \(eB\) region) and magnetization effects (dominant in the high \(eB\) region). The Tsallis parameter \(q\) further amplifies the above effects, especially most significantly near the transition region, highlighting the synergistic role of non-extensive statistics and chiral imbalance in shaping the pressure anisotropy of QCD matter under strong magnetic fields, while also confirming the amplifying effect of non-extensive statistics on the magnetic response of the system.
	
As shown in Fig.~\ref{fig:8}, the squared speed of sound \(c_s^2\) exhibits a pronounced dip near the critical temperature \(T_c\), and this dip shifts toward lower temperatures as the Tsallis parameter \(q\) increases. This dip reflects the softening of the equation of state in the transition region. When the magnetic field takes the value \(eB = 0.01\ \text{GeV}^2\), \(c_s^2\) overall exceeds the Stefan-Boltzmann (S-B) limit, indicating the presence of non-ideal behavior in the system. As the magnetic field further increases, for near-equilibrium cases (\(q = 1.001\) and \(q = 1.1\)), \(c_s^2\) rises steadily and gradually approaches the S-B limit. However, under strong non-equilibrium conditions (\(q = 1.2\)), \(c_s^2\) remains above the S-B limit in the high-temperature region, indicating that due to the dominance of non-extensive effects, the system continues to exhibit non-ideal behavior.
	
The influence of the chiral chemical potential \(\mu_{5}\) is also significant: a larger \(\mu_{5}\) increases the magnitude of \(c_s^2\) and modifies the position and depth of the dip. In particular, comparing subplots (a) and (b), it can be seen that increasing \(\mu_{5}\) from \(0.1\ \text{GeV}\) to \(0.2\ \text{GeV}\) leads to a more pronounced dip, and its position shifts toward lower temperatures, highlighting the regulatory role of chiral imbalance in the thermodynamic response near the phase transition. Nevertheless, the overall temperature dependence of \(c_s^2\) and the convergence properties (for the cases with \(q \leq 1.1\)) remain qualitatively unchanged, indicating that although \(\mu_{5}\) alters the detailed dynamical behavior of the phase transition process, it does not fundamentally change the evolution characteristics of \(c_s^2\).

	\section{Summaries and Conclusions}\label{sec:04 summary}

    This investigation of the two-flavor NJL model within Tsallis non-extensive statistics reveals that the non-equilibrium conditions inherent to heavy-ion collisions fundamentally reshape the QCD phase diagram and thermodynamic properties under strong magnetic fields and chiral imbalance. The Tsallis parameter $q$ acts as a powerful proxy for the critical off-equilibrium dynamics, moving beyond a mere phenomenological description to expose underlying physical mechanisms.
	
    Our results show that the critical temperature \(T_c\) decreases monotonically with increasing \(q\), meaning that non-equilibrium conditions can promote the restoration of chiral symmetry at lower temperatures. Regardless of which cutoff scheme is adopted, \(T_c\) decreases monotonically with increasing non-extensive parameter \(q\). In the range of \(q \approx 1.1\)–\(1.2\), the calculation results from the two schemes are in excellent agreement, and this self-consistency further supports the reasonableness of selecting \(q\) in the range of \(1.1\)–\(1.2\) in non-extensive statistical calculations.

    The chiral chemical potential \(\mu_5\) has a significant influence on the phase structure: as the magnetic field increases, \(T_c\) decreases significantly with increasing \(\mu_5\); at the same time, the value of \(\mu_5\) corresponding to the critical endpoint (CEP) also shifts significantly toward lower chemical potentials as \(q\) increases. In the case of fixed \(\mu_5\), the system exhibits an effect analogous to magnetic catalysis (MC); whereas when \(\mu_5\) takes the magnetic-field-dependent form \(\mu_5 = 0.5\sqrt{eB}\), especially in the weak-field region, the system exhibits characteristics analogous to inverse magnetic catalysis (IMC). Furthermore, when \(q > 1\), \(T_c\) exhibits non-monotonic behavior with respect to \(eB\), highlighting the complex coupling between the magnetic field and non-equilibrium effects.

    The core finding is the systematic reduction of the \(T_c\) with increasing \( q \). This signifies that the non-thermal fluctuations and long-range correlations encapsulated by $q > 1 $ provide an alternative pathway for chiral symmetry restoration, effectively lowering the energy barrier for melting the chiral condensate. The physical essence is that the system's departure from Boltzmann--Gibbs equilibrium accelerates its transition to a chirally symmetric phase, implying that the experimentally observed transition temperature may be a dynamic, non-equilibrium observable rather than a static equilibrium property.

	A clear pressure anisotropy is observed under strong magnetic fields. The longitudinal pressure $P_{\parallel}$ increases monotonically with $eB$, while the transverse pressure $P_{\perp}$ displays non-monotonic behavior, resulting from the competition between Landau quantization and magnetization. This anisotropy is further amplified by the Tsallis parameter $q$, especially near the transition between low- and high-field regimes. The squared speed of sound $c_s^2$ exhibits a dip near $T_c$, which shifts to lower temperatures as $q$ increases. For near-equilibrium cases ($q\approx1$), $c_s^2$ converges to the Stefan--Boltzmann limit at high temperatures, whereas under strong non-equilibrium ($q = 1.2 $), it exceeds this limit, reflecting persistent non-ideal behavior. The chiral chemical potential $\mu_{5}$ notably modifies the thermodynamic response: it enhances the sensitivity of pressure anisotropy to the magnetic field and alters the depth and position of the sound velocity dip. These results emphasize that chiral imbalance significantly interacts with magnetic and non-extensive effects, enriching the phase structure and thermodynamic properties of QCD matter.

	In summary, the Tsallis--NJL model provides a robust framework for exploring QCD phase transitions under non-equilibrium conditions. Our findings suggest that short-timescale dynamics in heavy-ion collisions can substantially lower the critical temperature and modify key thermodynamic observables. The results compellingly argue that the phase structure and thermodynamics of the QGP are co-determined by external conditions ($eB$, $\mu_{5}$) and the internal, non-equilibrium statistical state ($q$). This provides a critical theoretical lens for interpreting heavy-ion collision data, suggesting that the quest for the QCD phase diagram must account for the profound impact of the system's rapid, far-from-equilibrium evolution.
	
	\section*{Acknowledgments}
	This work was supported by the National Natural Science Foundation of China (Grants No. 12575144, and No. 11875178).
	
	\section*{References}
	
	\nocite{*}
	\bibliography{re4}
	
\end{document}